\documentstyle[12pt,psfig,epsf]{article}

\newcommand{\lesssim}{ \mathop{}_{\textstyle \sim}^{\textstyle <} }

\begin{document}
\begin{titlepage}
\begin{center}

\hfill    LBNL-40535\\
\hfill UCB-PTH-97/37 \\
\hfill July, 1997\\

\vskip .25in

{\large Determining $\tan\beta$ at the NLC with SUSY Higgs
Bosons\footnote{Talk presented by T. Moroi at the SUSY'97 Conference,
May 27-31, Philadelphia, PA, USA.}}

\vskip 0.3in

Jonathan L. Feng$^{ab}$ and Takeo Moroi$^a$

\vskip 0.15in

{{}$^a$ \em Theoretical Physics Group,
     Lawrence Berkeley National Laboratory\\
     University of California, Berkeley, CA 94720, U.S.A.}

\vskip 0.1in

{{}$^b$ \em Department of Physics\\
     University of California, Berkeley, CA 94720, U.S.A.}
        
\end{center}

\vskip .1in

\begin{abstract}

We examine the prospects for determining $\tan\beta$ from heavy Higgs
scalar production in the minimal supersymmetric standard model at a
future $e^+e^-$ collider.  Our analysis is independent of assumptions
of parameter unification, and we consider general radiative
corrections in the Higgs sector.  Bounds are presented for $\sqrt{s} =
500$ GeV and 1 TeV, several Higgs masses, and a variety of integrated
luminosities. For all cases considered, it is possible to distinguish
low, moderate, and high $\tan\beta$.  In addition, we find stringent
constraints for $3\lesssim\tan\beta\lesssim 10$, and, for some
scenarios, also interesting bounds on high $\tan\beta$ through
$tbH^{\pm}$ production. Such measurements may provide strong tests of
the Yukawa unifications in grand unified theories and make possible
highly precise determinations of soft SUSY breaking mass parameters.

\end{abstract}

\end{titlepage}

\renewcommand{\thepage}{\roman{page}}
\setcounter{page}{2}
\mbox{ }

\vskip 1in

\begin{center}
{\bf Disclaimer}
\end{center}
\vskip .2in
\begin{scriptsize}
\begin{quotation}
This document was prepared as an account of work sponsored by the
United States Government. While this document is believed to contain
correct information, neither the United States Government nor any
agency thereof, nor The Regents of the University of California, nor
any of their employees, makes any warranty, express or implied, or
assumes any legal liability or responsibility for the accuracy,
completeness, or usefulness of any information, apparatus, product, or
process disclosed, or represents that its use would not infringe
privately owned rights.  Reference herein to any specific commercial
products process, or service by its trade name, trademark,
manufacturer, or otherwise, does not necessarily constitute or imply
its endorsement, recommendation, or favoring by the United States
Government or any agency thereof, or The Regents of the University of
California.  The views and opinions of authors expressed herein do not
necessarily state or reflect those of the United States Government or
any agency thereof, or The Regents of the University of California.
\end{quotation}
\end{scriptsize}
\vskip 2in

\begin{center}
\begin{small}
{\it Lawrence Berkeley Laboratory is an equal opportunity employer.}
\end{small}
\end{center}
\newpage

\renewcommand{\thepage}{\arabic{page}}
\setcounter{page}{1}
\setcounter{footnote}{0}

\section{INTRODUCTION}
\label{sec:Introduction}

Supersymmetry (SUSY) is an attractive target of future high energy
experiments, and the discovery of supersymmetric particles is eagerly
anticipated at proposed new colliders. The discovery of superparticles
is, however, not the end of the story. Once SUSY is found, we will be
at a new stage in high energy physics. First of all, we should check
the supersymmetric relations among various parameters to really
confirm supersymmetry. We can also begin the exciting investigation of
physics of high energy scales by using renormalization group analysis:
the SUSY breaking parameters as well as the coupling constants contain
information about high energy scales, and they can give us some hints
about new physics, such as grand unification (GUT), flavor symmetry,
or the mechanism of SUSY breaking. For these programs, accurate
determinations of the parameters in the lagrangian are crucial and
essential.

Among the various parameters in SUSY models, $\tan\beta$ is one of the
most important. One reason is that $\tan\beta$ plays a significant
role in relating experimental observables to the parameters in the
lagrangian. For example, reconstructions of the Yukawa coupling
constants, mass matrices of the charginos and neutralinos, and the
SUSY breaking masses of sfermions require a knowledge of
$\tan\beta$. Furthermore, a precise determination of $\tan\beta$ can
be a check of various models that prefer a specific value of
$\tan\beta$.

In this study, we discuss the prospects for $\tan\beta$ determination
from the production and decay of Higgs bosons at the Next $e^+e^-$
Linear Collider (NLC)~\cite{JLC,NLC}. Detailed study of Higgs
properties can give us a good determination of $\tan\beta$, since the
interactions of the (heavy) Higgs bosons with quarks and leptons
depend on $\tan\beta$. This possibility has been studied in a model
independent approach~\cite{hep-ph/9612333} and in the framework of the
minimal supergravity~\cite{hep-ph/9610495}.  The NLC will be a good
place for the detailed study of Higgs properties, and $\sim 1000$
heavy Higgs pairs can be produced if the heavy Higgs bosons are
kinematically accessible. By using the standard collider parameters
($\sqrt{s}=500{\rm ~GeV}$ and ${\cal L}=50{\rm ~fb^{-1}/yr}$ for
phase~I, and $\sqrt{s}=1{\rm ~TeV}$ and ${\cal L}=200{\rm
~fb^{-1}/yr}$ for phase~II), we estimate the expected accuracy of the
measured $\tan\beta$ as a function of the actual value of $\tan\beta$.
We will see that the error can be as small as $O(10~\%)$ or less for
most of the theoretically interesting $\tan\beta$ values.

In determining $\tan\beta$ with Higgs bosons, the model dependence is
weak, {\it i.e.}, to determine $\tan\beta$ from Higgs bosons, we
require only that (heavy) Higgs bosons be produced at the NLC. In
order to make this point clear, we do not make any assumption that
strongly depends on some specific model, such as the minimal
supergravity model. In our analysis, we assume only that nature is
described by the field content of the minimal supersymmetric standard
model (MSSM). All the relevant parameters which are needed to
determine $\tan\beta$ are then required to be measured experimentally.
Furthermore, we emphasize that our approach results in quite an
accurate measurement of $\tan\beta$ if the actual $\tan\beta$ is in
the moderate region ($\sim$ 3 -- 10). Several other methods have been
proposed to determine $\tan\beta$, such as those using
charginos~\cite{PRD52-1418}, staus~\cite{PRD54-6756}, or the muon
$(g-2)$~\cite{PRD53-6565}.  The discovery of
$H,A\rightarrow\tau\bar{\tau}$ at the LHC may also be used to set a
lower bound on $\tan\beta$~\cite{LHC}, though, in general, the study
of the heavy Higgs sector appears to be one of the most challenging
for the LHC~\cite{hep-ph/9602238}.  However, all of these methods do
not give good results if the underlying $\tan\beta$ is in the moderate
region.  Thus, our method is complementary to the other analyses.

\section{HIGGS BOSONS IN THE MSSM}

First, let us briefly review the Higgs sector in the
MSSM~\cite{HHG}. The MSSM contains two Higgs doublets:
 \begin{eqnarray}
  H_1 = 
   \left(\begin{array}{c} H_1^0 \\ H_1^- \end{array}\right),
  ~~~
  H_2 = 
   \left(\begin{array}{c} H_2^+ \\ H_2^0 \end{array}\right).
 \end{eqnarray}
 When the neutral components of these Higgs fields obtain vacuum
expectation values (VEVs), electroweak symmetry is broken. One
combination of VEVs is constrained so that we obtain the correct
value of the Fermi constant: $2(\langle H_1^0\rangle^2 +\langle
H_2^0\rangle^2)\equiv v^2\simeq (246{\rm ~GeV})^2$. On the other hand,
their ratio is the free parameter which is $\tan\beta$:
 \begin{eqnarray}
  \tan\beta = \langle H_2^0\rangle / \langle H_1^0\rangle .
 \end{eqnarray}

By expanding the Higgs fields around their VEVs, we obtain physical
Higgses as well as the Nambu-Goldstone bosons. In order to obtain the
physical modes, it is more convenient to use another basis $\Phi_1$ and
$\Phi_2$:
 \begin{eqnarray}
  \left(\begin{array}{c} \Phi_1 \\ \Phi_2 \end{array}\right)
  \equiv
  \left(\begin{array}{cc} \cos\beta & \sin\beta \\
                          -\sin\beta & \cos\beta
  \end{array}\right)
  \left(\begin{array}{c} H_1 \\ i\sigma^2 H_2^* 
  \end{array}\right).
 \end{eqnarray}
 In this basis, $\Phi_1$ gets a VEV, while $\Phi_2$ does not. We expand
$\Phi_1$ and $\Phi_2$ as
 \begin{eqnarray}
  \Phi_1 &=&
  \left(\begin{array}{c} (v + \phi_1 + iG^0)/\sqrt{2} \\ 
                         G^- \end{array}\right),
 \\
  \Phi_2 &=&
  \left(\begin{array}{c} (\phi_2 + iA)/\sqrt{2} \\ 
                         H^- \end{array}\right).
 \end{eqnarray}
 Then, from the fact that electroweak symmetry is broken by the VEV of
$\Phi_1$, the Nambu-Goldstone bosons $G^0$ and $G^\pm$ are contained
only in $\Phi_1$. The other fields ($\phi_1$, $\phi_2$, $A$, and
$H^\pm$) are physical degrees of freedom. The pseudoscalar $A$ and charged
Higgs $H^\pm$ are mass eigenstates. On the other hand, the CP-even
scalars, $\phi_1$ and $\phi_2$, mix in the mass matrix.  Mass
eigenstates, $h$ and $H$, can be obtained by a unitary transformation:
 \begin{eqnarray}
  \left(\begin{array}{c} h \\ H \end{array}\right)
  \equiv
  \left(\begin{array}{cc} 
   \sin(\beta-\alpha) & \cos(\beta-\alpha) \\
   - \cos(\beta-\alpha) & \sin(\beta-\alpha)
  \end{array}\right)
  \left(\begin{array}{c} \phi_1 \\ \phi_2
  \end{array}\right),
 \end{eqnarray}
 where the unitary matrix is parametrized by a new parameter $\alpha$.
We define $h$ to be lighter than $H$. 

{\it At tree level}, mass and mixing parameters are related to each
other; once we fix $m_A$ and $\tan\beta$, all the masses and the
mixing parameter $\alpha$ are fixed.  However, the tree level
relations can be significantly modified by radiative
corrections~\cite{1loop}, and hence it may be dangerous to assume tree
level relations in the analysis. Therefore, we regard all the masses
and mixings as parameters to be measured by experiments, and
uncertainties in these measurements enter our analysis as systematic
errors.

Here, we comment on the so-called ``decoupling limit'' of the heavy
Higgses. When $m_A$ is much larger than $m_Z$, the mixing between
$\Phi_1$ and $\Phi_2$ becomes small: $\cos(\beta -\alpha)\rightarrow
0$. In this limit, $\Phi_1$ behaves like the standard model Higgs,
while the heavy Higgses ($H$, $A$, and $H^\pm$) is like an extra
doublet with degenerate mass. In our study, we assume that the charged
Higgs mass is heavier than the top quark mass, which implies that the
decoupling limit is more or less realized.

In the decoupling limit, $h$ is mainly produced in association with
the $Z$ boson ($e^+e^-\rightarrow Zh$), while for the heavy Higgses,
pair productions ($e^+e^-\rightarrow AH$, $e^+e^-\rightarrow H^+H^-$)
are the most important processes. The cross section for $H^+H^-$
production is independent of $\alpha$ and $\beta$, while those for
$Zh$ and $AH$ are both proportional to $\sin^2(\beta -\alpha)$. From
the precise measurement of the cross section of the process
$e^+e^-\rightarrow Zh$, $\sin^2(\beta -\alpha)$ is well determined
with accuracy $\sim 2~\%$~\cite{JLC}. Therefore, the cross sections
can be estimated with small errors in this study. Notice that the
cross sections for other processes ($e^+e^-\rightarrow ZH$,
$e^+e^-\rightarrow Ah$) are also calculable, but they are too
suppressed to be important since they are proportional to
$\cos^2(\beta -\alpha)$.

The Higgs bosons $H_1$ and $H_2$ are responsible also for
fermion masses. They have interactions of the form
 \begin{eqnarray}
  L_{\rm Y} = y_t H_2 q_L t_R^c + y_b H_1 q_L b_R^c 
  + y_\tau H_1 l_L \tau_R^c,
 \label{Yukawa}
 \end{eqnarray}
 where the $y_t$, $y_b$, and $y_\tau$ terms are the Yukawa couplings
for $m_t$, $m_b$, and $m_\tau$, respectively\footnote{The interactions
of the Higgs bosons with the first and second generations are too
suppressed to be important in our analysis, and we neglect them.}. By
substituting the mass eigenstates into $H_1$ and $H_2$, we obtain the
interactions of the physical Higgs bosons with the fermions. As noted
before, $h$ behaves like the standard model Higgs in the decoupling
limit, so its interactions are insensitive to $\tan\beta$. On the
other hand, the interactions of the heavy Higgses ($H$, $A$, and
$H^\pm$) with fermions strongly depend on $\tan\beta$. For example,
the coupling of $H^\pm$ to $t_R^c$ and $b_L$ is proportional to
$\cot\beta$, while that to $t_L$ and $b_R^c$ and that to $\tau_R$ and
$\nu_\tau$ are proportional to $\tan\beta$. Thus, if $\tan\beta$ is
small, charged Higgs bosons mainly decay into top and bottom quarks,
but the branching ratio of $H^\pm$ decaying into $\tau$ and $\nu_\tau$
increases as $\tan\beta$ gets large. Similarly, branching ratios of
$H$ and $A$ are also sensitive to $\tan\beta$\footnote{In fact, the
Yukawa couplings of $H$ depend on $\alpha$. In our numerical study, we
include this correction. The $\alpha$ dependence of the interaction
becomes weak in the decoupling limit.}.  Therefore, if we measure the
branching ratio of the heavy Higgs bosons, we can constrain
$\tan\beta$.

\section{EXPERIMENTAL SIMULATION}

In this section, we present our basic idea for determining
$\tan\beta$.  As stated in the previous section, the interactions of
the heavy Higgs bosons depend on $\tan\beta$, and hence the branching
ratios of the Higgses are sensitive to $\tan\beta$. This implies that
measurements of cross sections for heavy Higgs production
with various final states can give us some information about
$\tan\beta$.

For this purpose, we use the following observations. First of all,
Higgs bosons mostly decay into particles in the third generation.
Thus, a large number of $b$-jets is expected in Higgs production
events if the Higgses decay into hadrons. In the NLC, $b$-jets are
expected to be selected with a high $b$-tagging efficiency
($\epsilon_b\sim 60~\%$)~\cite{hep-ph/9703330}, so this can
effectively reduce the background. Furthermore, if the charged Higgs
decays into $\tau\nu_\tau$, a single high energy lepton (or energetic
hadrons with very low multiplicity) is expected as a decay product of
$\tau$. This can be a striking signal of charged Higgs production
followed by leptonic decay. The last point concerns the large
$\tan\beta$ limit. If $\tan\beta$ becomes larger than $\sim$ 10, all
the branching ratios of the heavy Higgses lose their sensitivities to
$\tan\beta$, and $\tan\beta$ cannot be constrained well from pair
production processes. However, in this case, the cross section of the
process $e^+e^-\rightarrow tbH^\pm$ can be enhanced enough to be
observed, since the $tbH^\pm$ vertex is proportional to $\tan\beta$ in
the large $\tan\beta$ limit. Thus, in this region, the $tbH^\pm$
production process becomes useful.

Based on the above arguments, we use the following types of channels
in our analysis:
 \begin{itemize}
  \item[{\bf 1}~:~] $2b+l+q{\rm 's} +$kinematical cuts to select the
   ``$H^+H^-$'' mode.
  \item[{\bf 2}~:~] $2b+l+q{\rm 's} +$kinematical cuts to select the
   ``$tbH^{\pm}$'' mode.
  \item[{\bf 3}~:~] $3b+1l\ (+q{\rm 's})$.
  \item[{\bf 4}~:~] $3b+0,2,3,\ldots\ l\ (+q{\rm 's})$.
  \item[{\bf 5}~:~] $4b$.
  \item[{\bf 6}~:~] $4b+1l\ (+q{\rm 's})$.
  \item[{\bf 7}~:~] $4b+ 0,2,3,\ldots\ l\  (+q{\rm 's})$ (but not $4b$). 
  \item[{\bf 8}~:~] $5b\ (+\ l+q{\rm 's})$.
 \end{itemize} 
 In this list, ``$b$'' and ``$q$'' denote hadronic jets with and
without a $b$-tag, respectively, ``$l$'' denotes an isolated,
energetic $e$, $\mu$, or $\tau$, and particles enclosed in parentheses
are optional. In our analysis, we assume that hadronically-decaying
$\tau$ leptons may be identified as leptons, ignoring the slight
degradation in statistics from multi-prong $\tau$ decays.

Channel~1 is intended to select charged Higgs pair production events
with $tb\tau\nu_\tau$ final states, while channel~2 is designed to
isolate $tbH^\pm$ production with $H^\pm\rightarrow \tau\nu_\tau$.
These two channels have the same event topology. Furthermore, top
quark pair production may contribute to these channels as a
significant background, if followed by the decays $t\bar{t}\rightarrow
(bW^+)(\bar{b}W^-)\rightarrow (bl\nu_l)(\bar{b}qq')$. It is crucial to
distinguish these processes, so we impose certain sets of kinematical
cuts for this purpose. In imposing the kinematical cuts, the important
point is that, in these events (channels~1, 2 and the background), one
top quark decays only into hadrons, so we can reconstruct the top
quark system from the hadrons after first judiciously choosing one of
the two $b$-jets~\cite{hep-ph/9612333}.  In the $t\bar{t}$ pair
production case, the energy of the top quark is, in principle, equal
to the beam energy, while for channels~1 and 2, it tends to be smaller
than the beam energy.  Thus, we demand the reconstructed top energy to
be well below the beam energy to eliminate the $t\bar{t}$
background. In order to distinguish channels~1 and 2, we check the
total energy of the hadrons. In channel~1, all the hadrons are the
decay products of one charged Higgs, and hence the the total energy of
the hadrons is equal to the beam energy. On the other hand, in
channel~2, the total energy of the hadrons is likely to be larger than
the beam energy. Based on these two observations, we impose
kinematical cuts on channels~1 and 2. Such kinematical cuts are highly
effective once the relevant cut parameters are optimized, and
channels~1 and 2 can be well separated.  (For details, see
Ref.~\cite{hep-ph/9612333}.)

Channels~3 -- 8 receive contributions mainly from Higgs pair
production events followed by hadronic decays of the Higgses. In these
channels, we choose only events with large numbers of $b$-jets.  The
resulting backgrounds have been calculated in
Ref.~\cite{hep-ph/9612333} and are found to be sufficiently
suppressed. Therefore, we do not impose any kinematical cuts on these
channels.

Once the relevant channels are chosen, we can quantitatively estimate
the accuracy of the $\tan\beta$ determination from measurements of the
cross sections of these channels. For this purpose, we must first
choose the relevant underlying parameters that fix the Higgs
potential. In our analysis, we used a Higgs potential with only the
leading $m_t^4$ correction from the top-stop loop, with stop mass
1~TeV. All the masses and mixings in the Higgs sector are then fixed
if we determine the charged Higgs mass and $\tan\beta$.  Notice that
we use the simple form of the Higgs potential only to generate the
event samples, and do not use any theoretical assumptions in the
actual determination of $\tan\beta$.  It is therefore straightforward
to extend our analysis to the case with general radiative corrections.

With the physical underlying parameters, we estimate the cross
sections in channels~1 -- 8. These cross sections then determine the
number of events that would actually be observed in each channel. We
then postulate a hypothetical $\tan\beta$, calculate the cross
sections in channels~1 -- 8 based on the postulated $\tan\beta$, and
check whether the postulated $\tan\beta$ is consistent with
observations. To quantify the argument, we define 
\begin{eqnarray}
\Delta \chi^2 = \sum_{i : {\rm channel}} \frac{(N_i - N'_i)^2}
{{\sigma^i}_{\rm stat}^2 + {\sigma^i}_{\rm syst}^2},
\end{eqnarray}
 where $N_i$ ($N_i'$) is the number of events in channel~$i$ for the
underlying (postulated) $\tan\beta$, and the quantities $\sigma_{\rm
stat}^i$ and $\sigma_{\rm syst}^i$ are the statistical and systematic
errors for channel $i$, respectively. For simplicity, we add
$\sigma_{\rm stat}^i$ and $\sigma_{\rm syst}^i$ in quadrature.
The number of events has two origins: the signal and the background.
Notice that the background cross section is independent of
$\tan\beta$.  Therefore, if some channel is dominated by the
background, it cannot give a significant contribution to
$\Delta\chi^2$.

The statistical error is $\sigma_{\rm stat}^i = \sqrt{N'_i}$, while
the systematic error is given by \begin{equation} {\sigma^i}_{\rm
syst}^2 = \sum_P \left( \frac{\partial N'_i}{\partial P} \Delta P
\right)^2, \end{equation} where the sum is over all quantities $P$
that enter in the calculation of the numbers of events, and which
therefore contribute systematic uncertainties.  We include
uncertainties from the following quantities (the uncertainties we use
are in parentheses): the bottom quark mass (150 MeV~\cite{JLC}),
$\cos^2(\beta -\alpha)$ (2 \%~\cite{JLC}), the hadronic decay width of
the heavy Higgses (20 \%~\cite{SUSYQCD}), $Br(H\rightarrow hh)$ (10
\%~\cite{PLB375-203}), the $b$-tagging efficiency (2 \%~\cite{Jackson}),
and heavy Higgs masses ($16{\rm ~GeV}/\sqrt{0.035N_H}$, where $N_H$ is
the number of Higgs pair production events~\cite{ZPC65-449}).

In the following, we will use $\Delta\chi^2$ to estimate the accuracy
of the $\tan\beta$ measurement. We will show contours of constant
$\Delta\chi^2=3.84$, which we refer to as 95~\% C.L. contours.

\section{NUMERICAL RESULTS}

In this section, we present a quantitative estimate of the expected
uncertainty of the $\tan\beta$ measurement at the NLC. Here, we assume
the absence of SUSY decays modes, whose effects are discussed in the
next section.

\begin{figure}
 \centerline{\psfig{file=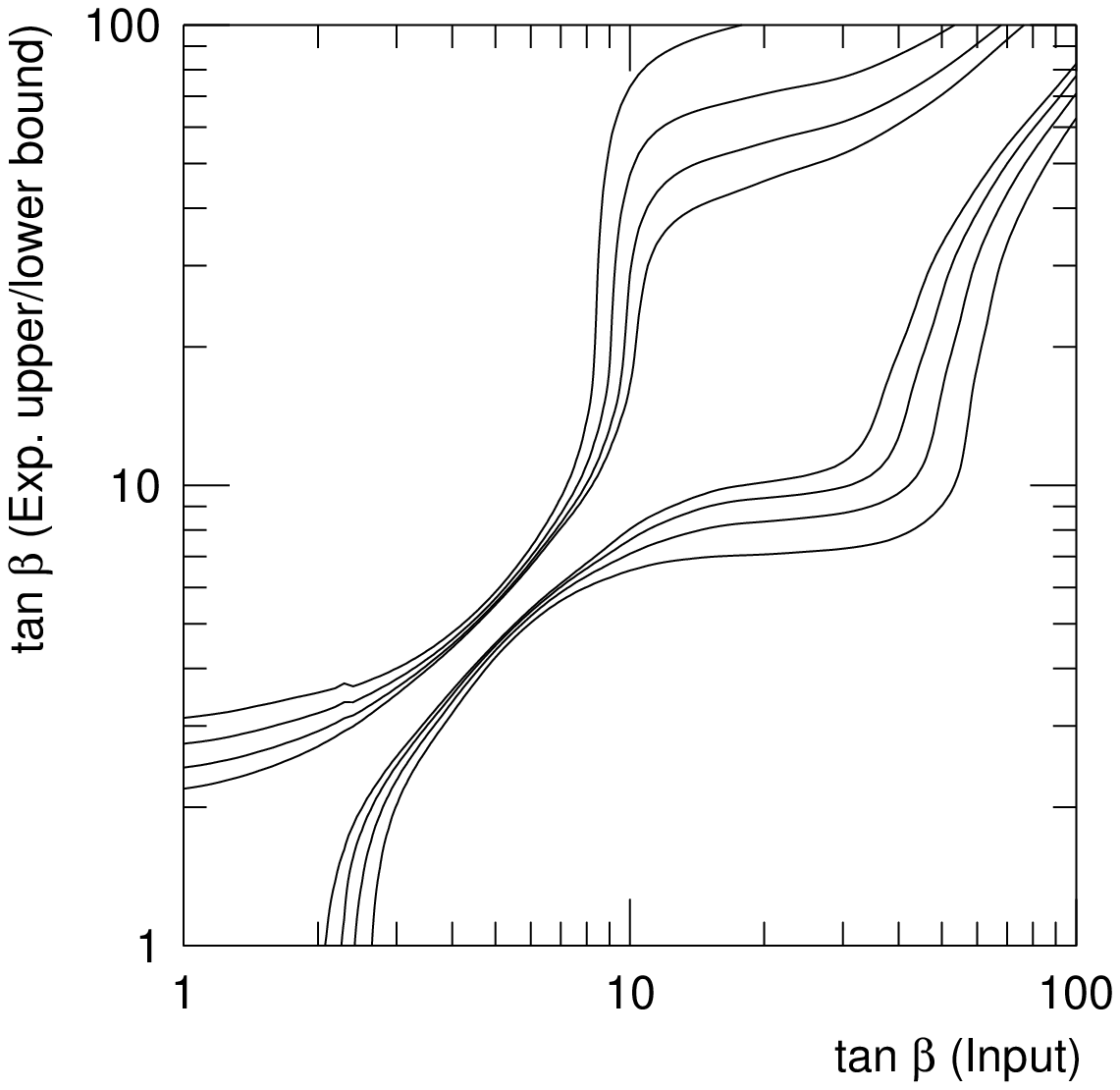,width=0.475\textwidth}}
 \caption{Accuracy of the measured $\tan\beta$ (95 \% C.L.)  for
$\protect\sqrt{s}=500$ GeV, $m_{H^{\pm}} = 200$ GeV, $\epsilon_b=60$
\%, and four integrated luminosities: 25, 50, 100, and 200 fb$^{-1}$
(from outside to inside).}
 \label{fig:m2ph1}
 \end{figure}

We start with the NLC with $\sqrt{s}=500$ GeV, and the charged Higgs
mass $m_{H^\pm}=200$ GeV. In this case, the decay of neutral Higgses
into $t\bar{t}$ pair is kinematically forbidden. The $\tan\beta$
dependence of the cross sections then comes mainly from charged Higgs
events. In Fig.~\ref{fig:m2ph1}, we plot the expected accuracy for
$\tan\beta$ as a function of the input value ({\it i.e.}, the actual
value) of $\tan\beta$. Here, we use four different integrated
luminosities: 25, 50, 100, and 200 fb$^{-1}$, which correspond to 0.5,
1, 2, and 4 years of running with design luminosity. The qualitative
behavior of the figure can be understood in the following way. If the
actual value of $\tan\beta$ is close to 1, the branching ratio of the
charged Higgs decaying into $\tau\nu_\tau$ is too suppressed to be
observed, and hence we may set an upper bound on $\tan\beta$ from the
non-observation of the $H^\pm$ leptonic decay mode.  If the underlying
$\tan\beta$ is in the moderate region ($\sim$ 3 -- 10), we observe
both the leptonic and hadronic decays of $H^\pm$. In this case, we can
determine $\tan\beta$ from these modes, as can be seen in
Fig.~\ref{fig:m2ph1}. If the actual $\tan\beta$ is larger than $\sim
10$, pair production of heavy Higgses allows us to set a lower limit
on $\tan\beta$. However, if $\tan\beta$ is large enough, the cross
section for the process $e^+e^-\rightarrow tbH^\pm$ (channel~2) can be
large enough to be observed, and both an upper and a lower limit on
$\tan\beta$ can be obtained from channel~2.

\begin{figure}
 \centerline{\psfig{file=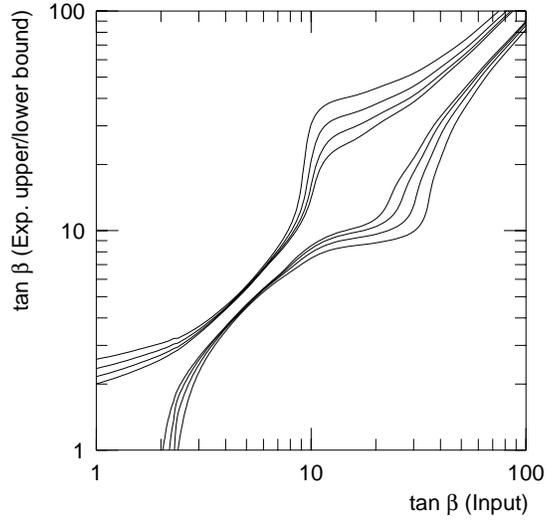,width=0.475\textwidth}}
 \caption{Accuracy of the measured $\tan\beta$ (95 \% C.L.)  for
$\protect\sqrt{s}=1$ TeV, $m_{H^{\pm}} = 200$ GeV, $\epsilon_b=60$ \%,
and four integrated luminosities: 100, 200, 400, and 800 fb$^{-1}$
(from outside to inside).}
 \label{fig:m2ph2}
 \end{figure}

\begin{figure}
 \centerline{\psfig{file=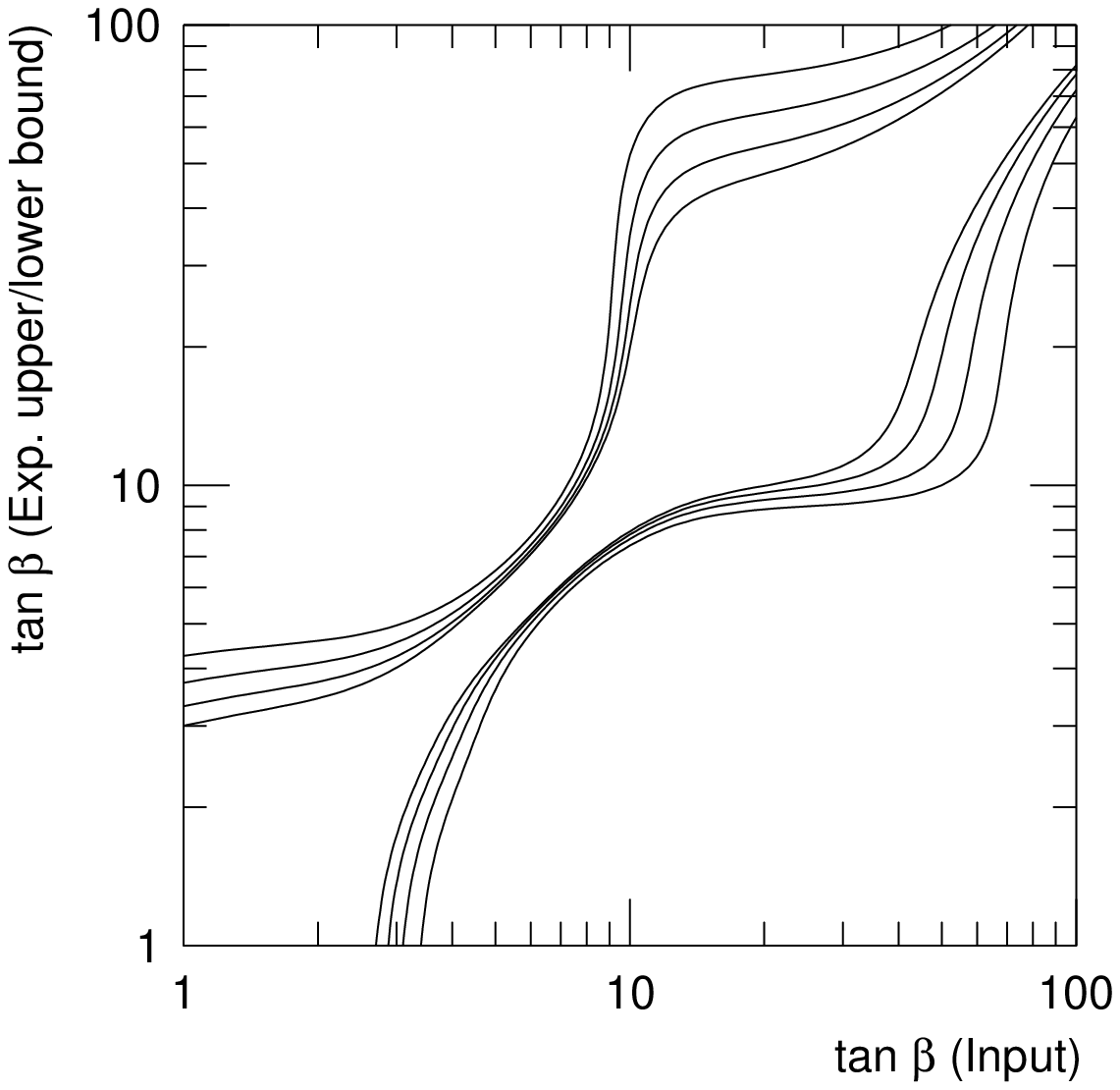,width=0.475\textwidth}}
 \caption{Same as Fig.~\protect\ref{fig:m2ph2}, except for
$m_{H^{\pm}} = 300$ GeV.}
 \label{fig:m3ph2}
 \end{figure}

\begin{figure}
 \centerline{\psfig{file=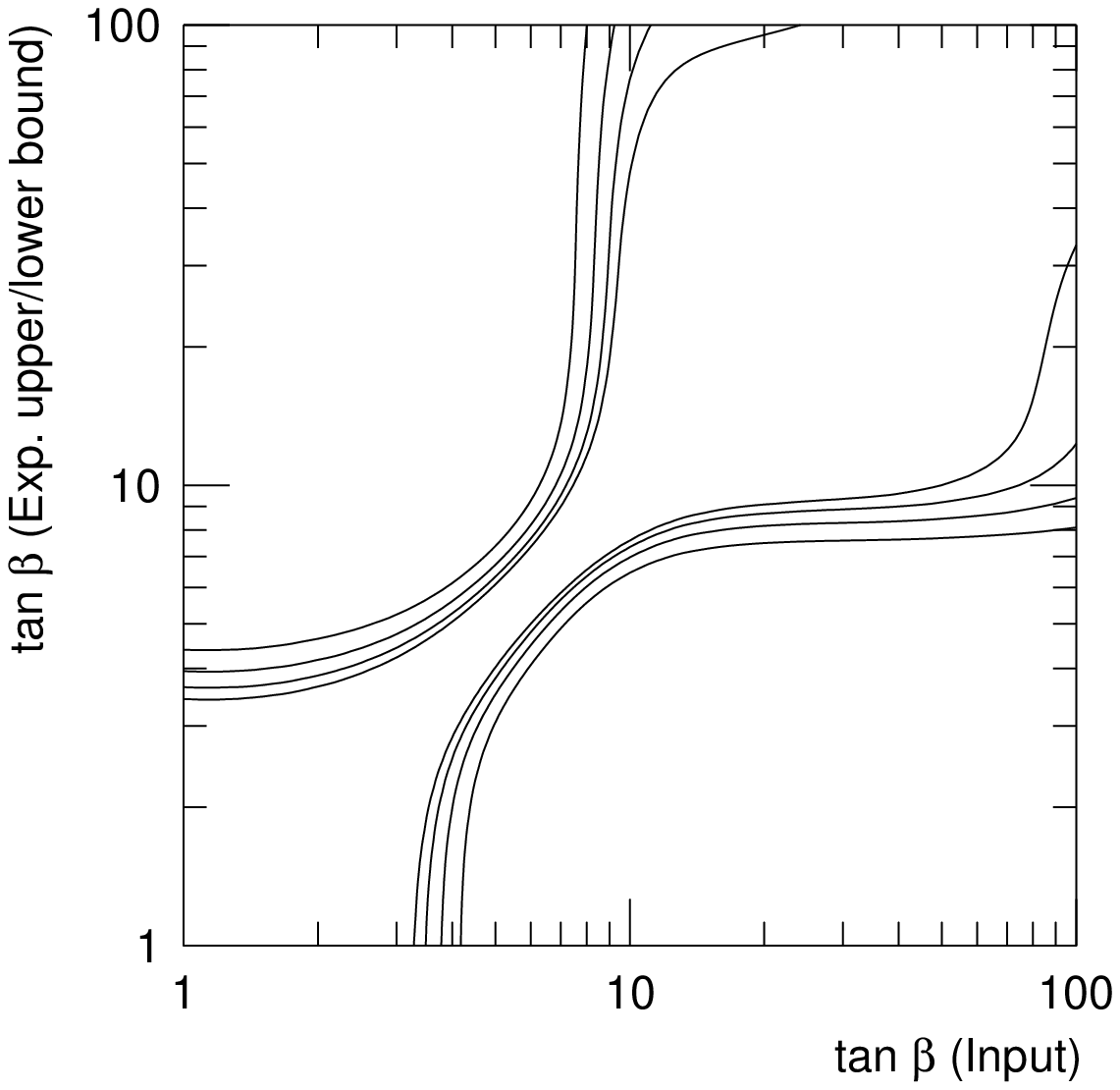,width=0.475\textwidth}}
 \caption{Same as Fig.~\protect\ref{fig:m2ph2}, except for
$m_{H^{\pm}} = 400$ GeV.}
 \label{fig:m4ph2}
 \end{figure}

If the heavy Higgses are not kinematically accessible, or even if they
are, it is advantageous to increase the beam energy.  Therefore, we
next present results for phase~II of the NLC with $\sqrt{s}=1$ TeV.
In Figs.~\ref{fig:m2ph2} -- \ref{fig:m4ph2}, we present results for
$m_{H^{\pm}} = $ 200 GeV, 300 GeV, and 400 GeV, respectively. For the
integrated luminosities, we use 100, 200, 400, and 800 fb$^{-1}$. In
particular, for $m_{H^{\pm}} = $ 200 GeV, we can see a great
improvement of the result, comparing Figs.~\ref{fig:m2ph1} and
\ref{fig:m2ph2}. For $m_{H^{\pm}} = $ 300 GeV, we can still expect a
precise determination of $\tan\beta$, especially if the underlying
$\tan\beta$ is moderate or large. For $m_{H^{\pm}} = $ 400 GeV, the
result is noticeably worse. There are mainly two reasons for
this. First, $\sqrt{s}$ is close to the threshold energy in this
case. The Higgs production cross sections then becomes small due to
the phase space suppression, and the statistical errors become
large. Furthermore, as the charged Higgs mass gets larger, the phase
space suppression of the process $H^\pm\rightarrow tb$ becomes less
significant, and the decay mode $H^\pm\rightarrow\tau\nu_\tau$ is
relatively suppressed. As a result, the branching ratio of the charged
Higgs loses its sensitivity to $\tan\beta$, contrary to the cases with
$m_{H^{\pm}} = $ 200 and 300 GeV. In fact, in the case with
$m_{H^{\pm}} = $ 400 GeV, $\tan\beta$ is mainly constrained by the
$AH$ production process with $4b$, $2b2t$, and $4t$ final
states. Note, however, that even in this case, we may distinguish low,
moderate, and high $\tan\beta$, and we can still obtain accurate
measurements of $\tan\beta$ if the underlying value of $\tan\beta$ is
in the moderate region.

Before closing this section, we comment on the effects of the
supersymmetric decay modes. Up to now, we have assumed that the Higgs
bosons decay only into standard model particles.  However, especially
when the Higgs masses are large, Higgses may decay into
superparticles, which one might think would lead to new sources of
large systematic errors.  Here, we would like to point out that {\it
this is not necessarily the case}.  Let us start with the decay mode
$H\rightarrow\tilde{l}_R\tilde{l}_R^*$. We can predict the branching
ratio for this mode, since this process is induced by $D$-term
interactions\footnote{Slepton masses will be measured at the NLC with
good accuracy.}. Thus, we only have to include this mode into the fit,
and no new large systematic uncertainties arise. The primary effect of
this decay mode is then to reduce the number of events from $AH$
production, and in fact, the numerical results are not changed
much~\cite{hep-ph/9612333}. The same argument applies to the decay of
$H$ into pairs of left-chirality sfermion. However, if the heavy
Higgses can decay into sfermion pairs with different chirality, such
as $H\rightarrow\tilde{l}_R\tilde{l}_L^*$, the branching ratios of the
heavy Higgses depend on additional MSSM parameters, such as $\mu$ and
the trilinear $A$ terms. These decay modes may then be a source of a
large systematic errors. Of course, the new parameters may also be
measured from different observables; for example, $\mu$ may be
measured from chargino and neutralino masses, and the trilinear scalar
couplings may be measured from left-right mixings.  A complete
analysis would therefore require a simultaneous fit to all of these
parameters.

Finally, we briefly consider decays to charginos and neutralinos.
These decay may be dominant in some regions of parameter space.
However, if only decays to the lighter two neutralinos and the lighter
chargino are available, and these are either all gaugino-like or all
Higgsino-like, as is often the case, these decays are suppressed by
mixing angles.  If we are in the mixed region, these decay rates may
be large, but in this case, all six charginos and neutralinos should
be produced, and the phenomenology is quite rich and complicated.

\section{DISCUSSION}

The results presented in the previous section indicate that
$\tan\beta$ may be well determined from the study of production and
decay of the heavy Higgses at the NLC. The uncertainty in the measured
$\tan\beta$ can be $O(10~\%)$ or less, depending on the underlying
parameters. For example, for $\sqrt{s}=500$ GeV, $m_{H^{\pm}} = 200$
GeV, and ${\cal L}=100$ fb$^{-1}$, the observed value $\tan\beta_{\rm
obs}$ will be in the following ranges for various input values of
$\tan\beta$:
\begin{eqnarray} \tan\beta ~{\rm (input)} = 2: & \tan\beta_{\rm obs} <
2.9, \nonumber \\ \tan\beta ~{\rm (input)} = 3: & 2.5 < \tan\beta_{\rm
obs} < 3.6, \nonumber \\ \tan\beta ~{\rm (input)} = 5: & 4.5 <
\tan\beta_{\rm obs} < 5.5, \nonumber \\ \tan\beta ~{\rm (input)} = 10:
& 7.6 < \tan\beta_{\rm obs} < 30, \nonumber \\ \tan\beta ~{\rm
(input)} = 60: & 40 < \tan\beta_{\rm obs} < 90. \nonumber
\end{eqnarray}

In our study, we have not adopted any assumption which strongly
depends on a specific model. Thus, if the heavy Higgses are
kinematically accessible at the NLC, we can expect to obtain a
constraint on $\tan\beta$ from the detailed study of heavy Higgs
bosons.

Finally, we discuss implications of precise determinations of
$\tan\beta$. First of all, it should be emphasized that a
determination of $\tan\beta$ may help us understand physics at very
high energy scales. For example, the simple SO(10) GUT predicts large
values of $\tan\beta$~\cite{so(10)}. The unification of $m_b$-$m_\tau$
based on SU(5)-type GUTs is an another example. It prefers a value of
$\tan\beta$ that is very large or close to 1~\cite{b-tau}. Precise
determinations of $\tan\beta$ can be excellent tests of these
scenarios.

The parameter $\tan\beta$ is also important for the determination of
the SUSY breaking scalar masses. Neglecting mixings, the physical
masses of sfermions are the sum of soft SUSY breaking masses and the
$D$-term contribution which depends on $\tan\beta$. Thus, we must know
$\tan\beta$ to determine soft SUSY breaking mass parameters from the
physical masses of sfermions. If $\tan\beta$ is completely unknown,
soft SUSY breaking masses may have large uncertainties even if we can
measure the sfermion masses very accurately. Measurement of
$\tan\beta$ reduce this uncertainty. The important point is that the
$D$-term contribution is proportional to $\cos 2\beta$, and hence it
becomes insensitive to $\tan\beta$ once $\tan\beta$ is larger than 3
-- 4. Given our result, the uncertainty related to the $D$-term
contribution is smaller than the error from the sfermion mass
measurement, if $\tan\beta$ is larger than 3 -- 4.  The spectrum of
the soft SUSY breaking masses may contain information about the origin
of SUSY breaking~\cite{PTPS123-507} and the gauge and/or flavor
structure at high scales~\cite{scalarmass}.

Furthermore, if we combine the $\tan\beta$ determination with
measurements of the Higgs masses, we may be able to gain some
information about the Higgs potential. The Higgs potential is strongly
constrained at the tree level, but radiative corrections are quite
significant. In particular, the top squark plays an important role,
and the precise determination of $\tan\beta$ as well as the Higgs
masses may give us some constraint on top squark masses.

In summary, determination of the $\tan\beta$ parameter can potentially
give us rich information about physics at the high scale, including
the possibility of GUTs, high energy flavor structures, and the origin
of SUSY breaking.

\section*{ACKNOWLEDGEMENT}

This work was supported in part by the Director, Office of Energy
Research, Office of High Energy and Nuclear Physics, Division of High
Energy Physics of the U.S.  Department of Energy under Contract
DE-AC03-76SF00098 and in part by the NSF under grant PHY-95-14797.
J.L.F. is supported by a Miller Institute Research Fellowship.

\newpage

\end{document}